# TIMING SYSTEM OF SPRING-8 BOOSTER SYNCHROTRON

N. Hosoda, T. Ohshima, T. Aoki, T. Asaka, H. Ego, K. Fukami, Y. Kawashima, Y. Ohashi, T. Takashima, H. Yonehara, SPring-8, Hyogo 679-5198, Japan

*Abstract*

The timing system of SPring-8 booster synchrotron generates various timing signals concerning beam injection, acceleration from 1 GeV to 8 GeV and ejection. We have improved the timing system of the synchrotron giving it better stability and flexibility. This improvement results the other merits of advanced operations, for example, storing electron beam in the synchrotron, changing the injection cycle from 1 Hz to the slower frequency to increase the RF knock-out (RF-KO) operation period and ejecting the low energy beam during ramping up.

## 1 INTRODUCTION

The SPring-8 facility consists of a 1 GeV linac with a repetition rate of 60 Hz, a booster synchrotron and an 8 GeV storage ring. The timing system installed in the RF station named E-station in the storage ring generates a 508.58 MHz RF reference signal and a 1-cycle signal at the rate of 1 Hz. The 1-cycle signal includes the information of an RF bucket address in the storage ring. The timing system of the synchrotron receives these signals and regenerates many timing signals such as a gun trigger, pulse magnet triggers and ramping patterns. For a single-bunch beam operation of the storage ring, an RF-KO system installed in the synchrotron rejects the electrons in satellite RF buckets around the main bucket [1].

## 2 REQUIREMENTS

There are four requirements for the timing system in the synchrotron.

The first is to synchronize the 508.58 MHz RF phase between the synchrotron and the storage ring to control the beam filling of storage ring.

The second is to suppress the time jitters between the gun trigger signal of the linac and the RF signal of the synchrotron as small as possible less than 100 ps. The time width of a beam from the linac for a single-bunch beam operation is 1 ns, and the time width of an RF bucket of the synchrotron is about 2 ns. If the energy spread of a beam is $\Delta E/E \geq 1\%$ and the gun trigger timing changes over 100 ps, the portion of the beam is not captured by an RF bucket of the synchrotron and the beam intensity of the synchrotron fluctuates. During the single-bunch beam operation, the synchrotron radiation users request the uniform beam intensity of each bunch of the storage ring. To realize it, the beam intensity from the synchrotron must be stable.

The third is to prepare the function to inject multi-pulse beam from the linac into the synchrotron and to inject into an RF bucket of the storage ring in 1 cycle to shorten beam-filling time in the storage ring. The harmonic number of the synchrotron is 672 and that of the storage ring is 2436. While the RF bucket of the synchrotron turns around 29 times, the RF bucket of the storage ring turns around 8 times. The RF bucket addresses of the synchrotron relative to that of the storage ring are summarized in Table 1.

The fourth is to add the flexibility of the advanced operations of the synchrotron such as storing the electron beam in the synchrotron, changing the injection cycle from 1 Hz to a slower frequency to increase the RF-KO operation period and ejecting the low energy beam during ramping up.

The timing system already satisfies the former three requirements. However, to satisfy the former two requirements with the higher precision, we improved the RF low power system last year. In addition, to improve the function to inject multi-pulse beam and to add the flexibility, we improved the timing system this year.

Table 1: RF bucket addresses of the synchrotron relative to that of the storage ring. The delay counts of M for two 508MHz SUC's are mentioned in section 3.2.

| Synchrotron revolution | Synchrotron RF bucket address | Storage ring revolution | M for 508MHz SUC with N=672 | M for 508MHz SUC with N=2436×8 |
|---|---|---|---|---|
| 0 | 0 | 0 | 0 | 1 |
| 3 | 420 | 1 | 420 | 2436 |
| 7 | 168 | 2 | 168 | 4872 |
| 10 | 588 | 3 | 588 | 7308 |
| 14 | 336 | 4 | 336 | 9744 |
| 18 | 84 | 5 | 84 | 12180 |
| 21 | 504 | 6 | 504 | 14616 |
| 25 | 252 | 7 | 252 | 17052 |

## 3 THE SYSTEM

### 3.1 Optic Fiber Link

All timing signals are linked in the timing system of the synchrotron located at the injector control room. We use two kinds of optic fibers. One is the single-mode phase-stabilized optic fiber supplied by Sumitomo Electric Industries Ltd. [2]. This is used for the distributions of 508.58 MHz RF reference signal and precise timing signals such as triggers for an electron gun and fast beam monitors. The RF reference signal from the storage ring is controlled under a phase-locked loop system (PLL) [3]. In addition, the RF reference signal to the RF low power system employs the PLL, too. The other one is the multi-mode graded-index optic fiber with core/cladding size of 50/125 μm. This is used for the distributions of triggers for magnets and slow monitors. The measured time jitters of this optic fiber system with a fiber length of 250 m is 440 ps as a standard deviation.

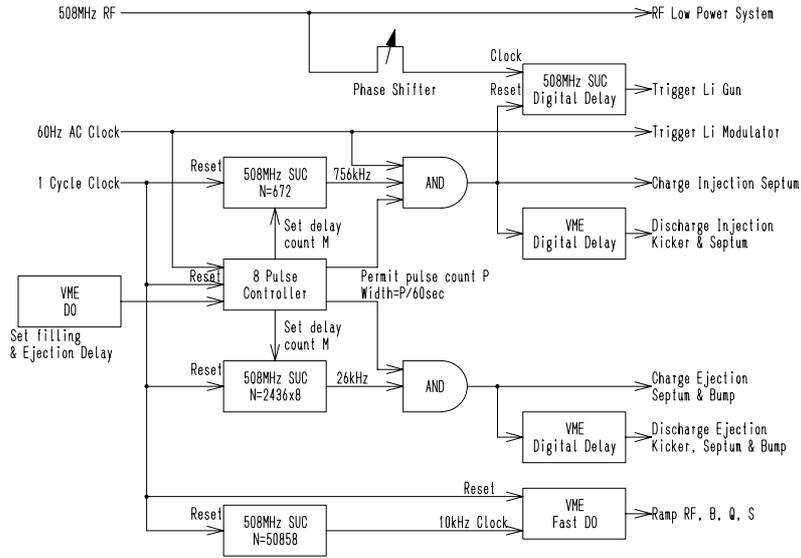

Fig. 1: Block diagram of the timing system of the synchrotron.

*3.2 508MHz SUC*

We can correctly count the RF bucket number using 508.58 MHz synchronous universal counters (508MHz SUC) [4]. The 508MHz SUC generates a clock with a frequency of 508.58/N MHz and a delay time of M/0.50858 ns from a counter reset signal. The N and M are positive integers.

Fig. 1 shows the block diagram of the timing system of the synchrotron. We use two 508MHz SUC's to control multi-pulse beam. One is to define the injection timing and is set N = 672. The other is to define the ejection timing and is set N = 2436×8 = 19488. The delay counts of M for two 508MHz SUC's are also summarized in Table 1. We set the beam filling of the synchrotron and the ejection timing to the 8-pulse controller as shown in Fig. 1 by using a VME digital out (DO) board. This controller controls the number of pulse and the delay count M's of two 508MHz SUC's for each pulse.

A 508MHz SUC is used for a gun trigger signal. If the trigger signal has some time jitters, the 508MHz SUC synchronizes the trigger signal with the RF reference again. Therefore, jitters smaller than 2 ns is completely cured. A phase shifter in front of this 508MHz SUC adjusts finely the timing of beam injection.

To generate the ramping patterns of the main magnets and the acceleration voltage (Vacc) using the VME fast DO [5], a 10 kHz clock is also generated by a 508MHz SUC.

*3.3 VME Digital Delay*

To discharge the pulse magnets such as kickers and septums for beam injection and ejection, we adopt VME 6 channel digital delay boards, Berkeley Nucleonics Corporation B951-2. The measured maximum time jitters at maximum delay of 167.8 ms is 0.9 ns. For multi-bunch beam operation, the time width of a beam from the linac is 40 ns and the kicker magnets have 60 ns flattop pulse width. The jitters is negligible.

*3.4 Phase Control of RF Signal*

The phase control part in the RF low power system of the synchrotron is shown in Fig. 2. We have two klystrons and each output power is provided into four 5-cell RF cavities. To make the beam acceptance larger, we control Vacc by changing the RF phase between two klystron outputs, keeping the RF voltage of the cavities constant.

The RF signals from the pick-up port of four cavities receiving the RF power from one klystron are combined. The combined signal is used as the input of the Cavity PLL. The combined RF signals are furthermore combined and total RF signal is used as the inputs of the Vacc PLL and the Vacc ALC (Auto Level Control feedback system). The Vacc PLL keeps the phase of total RF signal constant. The VME fast DO board generates the ramping pattern used as the reference signal of Vacc ALC. One of the Cavity PLL's controls a phase shifter using the Vacc ALC output as a reference and the other controls another phase shifter using the Vacc ALC output with reversed polarity. The relation of the phase between two klystron outputs and the Vacc during 1 cycle is shown in Fig. 3.

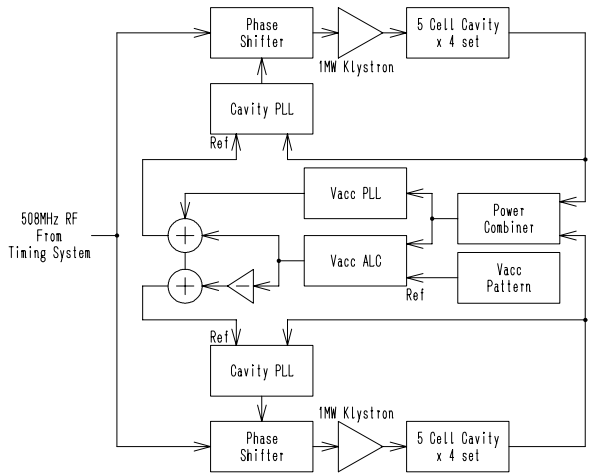

Fig. 2: Phase control part in the RF low power system of the synchrotron.

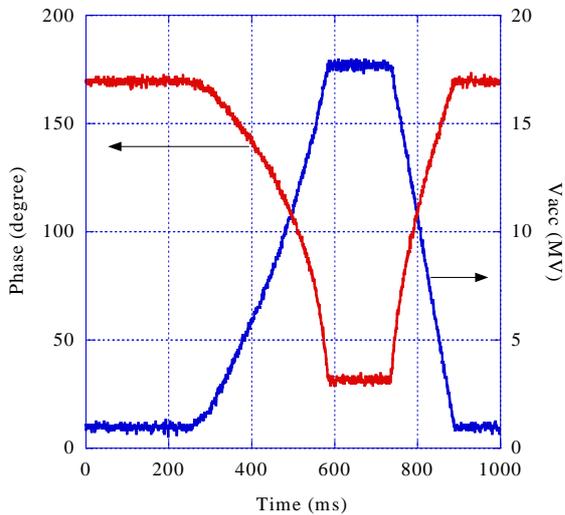

Fig. 3: Relation of the phase between two klystron outputs and the Vacc during 1 cycle.

## 4 PERFORMANCE

A fluctuation of the RF phase between the synchrotron and the storage ring is less than 0.3 degree by PLL's. We can correctly control the RF bucket position to inject a beam.

The measured time jitters of a gun trigger to the RF signal is 18 ps as a standard deviation. The intensity of the beam with 1 ns time width fluctuates about 5 %, mainly due to other causes.

Fig. 4 shows the electron beam current in the synchrotron during 1 cycle when eight pulses with 1 ns time width are injected and ejected. The Bending Magnet (BM) current is also drawn which represents the ramping pattern of the beam energy from 1 GeV to 8 GeV. We can control the beam intensity by changing the number of injected pulse to the synchrotron in eight steps.

The impurity of the single bunch in the storage ring is defined as the ratio of the number of electrons in the satellite RF buckets to the one in the main RF buckets. It is less than $10^{-9}$ order or under detection level. When the stored beam current decreased with time, we add the electrons to the same RF buckets. However, the impurity is not changed.

We started the study to eject the low energy beam during ramping up for low energy operation at the storage ring. We are planning to use the synchrotron as a storage ring. The DC operation of magnet power supplies is feasible up to the beam energy of 4.8 GeV. The functions to pause and to restart the ramping are already prepared.

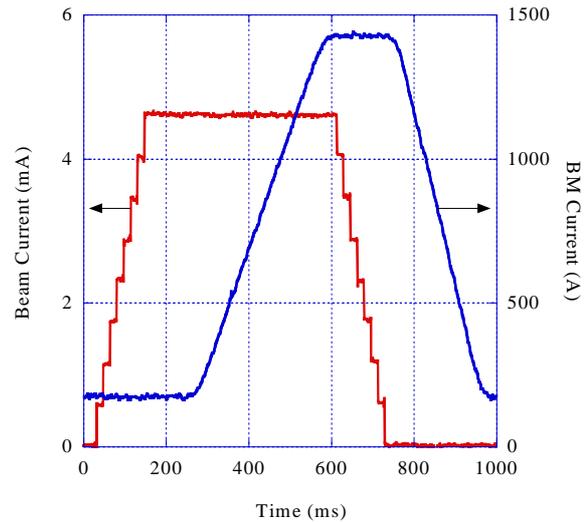

Fig. 4: Electron beam current in the synchrotron during 1 cycle when eight pulses with 1 ns time width are injected and are ejected. The BM current represents the ramping pattern of the beam energy from 1 GeV to 8 GeV.

## 5 CONCLUSION

The new timing system of the synchrotron has been constructed and confirmed to satisfy our requirements. The character of this system is the flexibility to meet the advanced operation of synchrotron.

## REFERENCE

[1] H. Suzuki *et al*., "Formation of a single-bunch beam in the booster synchrotron at SPring-8", Nucl. Instr. and Meth. A444, (2000) p.515.
[2] S. Tanaka *et al*., "Precise Timing Signal Transmission by a New Optical Fiber Cable", KEK Report 90-5 (1990).
[3] Y. Ohashi *et al*., "Performance of the 508MHz Reference and Timing Systems at SPring-8", ICALEPCS2001, San Jose, Nov. 2001.
[4] H. Suzuki *et al*., "508.58 MHz synchronous universal counter for beam control system of SPring-8", Nucl. Instr. and Meth. A431, (1999) p.294.
[5] N. Hosoda *et al*., "Integration of the Booster Synchrotron Control System to the SPring-8 Control System", ICALEPCS'99, Trieste, Oct. 1999.